\documentclass[
longbibliography,
showpacs,
floatfix,
aps, 
prl,
twocolumn,
superscriptaddress,
amssymb,
]{revtex4-2}

\usepackage{amsmath}
\usepackage{tabulary}
\usepackage{tabularx}
\usepackage{url}
\usepackage[breaklinks=true]{hyperref}

\usepackage{graphicx}
\usepackage{times}
\usepackage{wasysym}
\usepackage{amssymb}
\usepackage{latexsym}

\usepackage{dcolumn}
\usepackage{amsfonts}
\usepackage{bm}
\usepackage{epsfig}

\newcommand{\be}{\begin{equation}}
\newcommand{\ee}{\end{equation}}
\newcommand{\bea}{\begin{eqnarray}}
\newcommand{\eea}{\end{eqnarray}}

\newcommand{\req}[1]{Eq.\,(\ref{#1})}

\newcommand{\rfig}[1]{Fig.\,\ref{#1}}

\def\rh{R_H}
\def\rl{R_L}

\def\nua{\nu_\alpha}
\def\nub{\nu_\beta}

\def\bip{B_\parallel}

\def\ta{72.5^\circ}
\def\tb{73.4^\circ}
\def\tc{76.6^\circ}
\def\td{74.4^\circ}

\begin{document}
\title{High-order two-component fractional quantum Hall states around filling factor \textbf{$\nu = 1$}}
\author{E. Bell}
\affiliation{School of Physics and Astronomy, University of Minnesota, Minneapolis, Minnesota 55455, USA}
\author{K. W. Baldwin}
\affiliation{Department of Electrical Engineering, Princeton University, Princeton, New Jersey 08544, USA}
\author{L. N. Pfeiffer}
\affiliation{Department of Electrical Engineering, Princeton University, Princeton, New Jersey 08544, USA}
\author{K. W. West}
\affiliation{Department of Electrical Engineering, Princeton University, Princeton, New Jersey 08544, USA}
\author{M. A. Zudov}
\email[Corresponding author: ]{zudov001@umn.edu}
\affiliation{School of Physics and Astronomy, University of Minnesota, Minneapolis, Minnesota 55455, USA}
\received{\today}

\begin{abstract}
Two-component fractional quantum Hall (2C-FQH) states in electron bilayers have been known for decades, yet their experimental realization remained limited to low-order fractions.
Here we report on several families of high-order 2C-FQH states that emerge when an in-plane magnetic field drives a controlled monolayer-to-bilayer transition in an ultra-high-mobility GaAs quantum well.
These families of states proliferate symmetrically toward the filling factor $\nu = 1$, from both $\nu = 2/3$ and $\nu  = 4/3$, thereby respecting particle-hole symmetry. 
Surprisingly, many unbalanced states (with unequal layer fillings) are more robust than their parent balanced states, defying the expected hierarchy of Jain sequences. 
Our findings substantially expand the known landscape of 2C-FQH states, highlighting the unexpected richness of the bilayer quantum Hall regime and opening new routes for probing the interplay of symmetry, topology, and interactions in quantum Hall systems. 
\end{abstract}

\maketitle


The discovery of the integer quantum Hall effect \cite{klitzing:1980} and the subsequent observation of the fractional quantum Hall (FQH) effect \cite{tsui:1982} opened a rich field for exploring correlated electronic phases in two-dimensional electron systems (2DESs) subjected to perpendicular magnetic fields $B_\perp$, strong enough to produce Landau quantization. 
Within the composite-fermion (CF) framework \cite{jain:1989}, FQH states are understood as integer quantum Hall states of CFs, quasiparticles formed by attaching an even number  ($2p$) of flux quanta to each electron.
Because of such flux attachment, CFs experience reduced magnetic field $B_\perp^\star=B_\perp-2pn_e\phi_0$ ($n_e$ is the electron density and $\phi_0$ is the magnetic flux quantum) and fill $B_\perp^\star$-generated Landau levels, known as $\Lambda$-levels.
As a result, FQH states at fractional electron filling factors $\nu = h n_e/e B_\perp$, commonly known as Jain fractions, are related to integer CF filling factors $n = 1, 2,  3,...\,$ as
\be
\nu_n^{\pm}=\frac{n}{2pn \pm 1}\,.
\label{eq.jain}
\ee

In 2DESs with additional degrees of freedom, more exotic and complex states become possible. 
For example, the layer degree of freedom ($\alpha,\beta$) and electron interactions give rise to a whole new class of so-called two-component FQH (2C-FQH) states.
These states include an excitonic Bose condensate, formed by electrons in one layer and holes in another, which occurs at a total filling factor $\nu = 1$ (i.e., each layer is half full, $\nu_\alpha = \nu_\beta = 1/2$) \citep{kellogg:2004,tutuc:2004,eisenstein:2004,eisenstein:2014,liu:2017,li:2017} and
the FQH state at $\nu = 1/2$ (each layer is a quarter full) \citep{halperin:1983,suen:1992,eisenstein:1992,suen:1994,shabani:2013}.
In addition, the layer degree of freedom allows for 2C-FQH states at filling factors 
\be
\nu_{n_\alpha n_\beta}^{\pm} \equiv \nu_{n_\alpha}^{\pm} +  \nu_{n_\beta}^{\pm}\,,
\label{eq.2c}
\ee
where $n_\alpha$ and $n_\beta$ are CF filling factors in layers $\alpha$ and $\beta$.
For the simplest two-flux CFs ($2p=2$), and without inter-layer charge transfer (tunneling), one can naturally expect balanced fractions at $\nu^{+}_{11} = 1/3 + 1/3 = 2/3$, $\nu^{+}_{22} = 2/5 + 2/5 = 4/5\,,...  \,$, and $\nu^{-}_{22} = 2/3 + 2/3 = 4/3$, $\nu^{-}_{33} = 3/5 + 3/5 = 6/5\,,...\,$. 
However, if charge transfer between the layers is allowed, more complex (unbalanced) 2C-FQH states can form.

Until now, experimental observations of unbalanced 2C-FQH states have been limited to low orders, with layer CF filling factors differing by at most one, namely  $\nu^+_{12} = 1/3 + 2/5 = 11/15$, $\nu^+_{23} = 2/5 + 3/7 = 29/35$, and $\nu^-_{23} = 2/3 + 3/5 = 19/15$ \citep{manoharan:1997,liu:2015,liu:2019}.
In this work, we present experimental evidence for many higher-order 2C-FQH states, which come from several families and can have $n_\beta - n_\alpha$ as high as 3.
These states emerge under an in-plane magnetic field, which controllably drives a monolayer-to-bilayer transition by coupling to the orbital motion of electrons confined to a GaAs quantum well \citep{smrcka:1995}.
Remarkably, the 2C-FQH states occur at filling factors which are symmetric about $\nu = 1$, revealing particle-hole symmetry, a feature absent in a single-component regime. 
Even more unexpectedly, many unbalanced states, e.g., $\nu^+_{34} = 55/63$ ($\nu^-_{45} = 71/63$) are stronger than the lower-order balanced states at $\nu^+_{33} = 6/7$ ($\nu^-_{44} = 8/7$), contrary to the expected monotonic weakening of higher-order Jain fractions.
These results highlight a much richer 2C-FQH landscape than previously known and open new directions for exploring bilayer quantum Hall physics.

Our samples are $4\times 4$ mm squares, with eight indium contacts at the corners and the midsides, made of a new-generation GaAs quantum well 
\citep{chung:2021}.
The quantum well has a width of 49 nm, an electron density $n_e \approx 1 \times10^{15}$ m$^{-2}$, and a mobility $\mu \approx 4 \times10^3$ m$^2$/Vs.
The longitudinal resistance $\rl$ and the Hall resistance $\rh$ were measured at a base temperature of $T \approx 20$ mK using a four-terminal, low-frequency (a few Hz) lock-in technique.
The in-plane magnetic field was introduced by tilting the sample by angle $\theta$ with respect to the sample normal.
We emphasize that both the exceptional quality of our 2DES and the ability to precisely control the tilt angle are crucial for our findings.

\begin{figure*}[t] 
\includegraphics{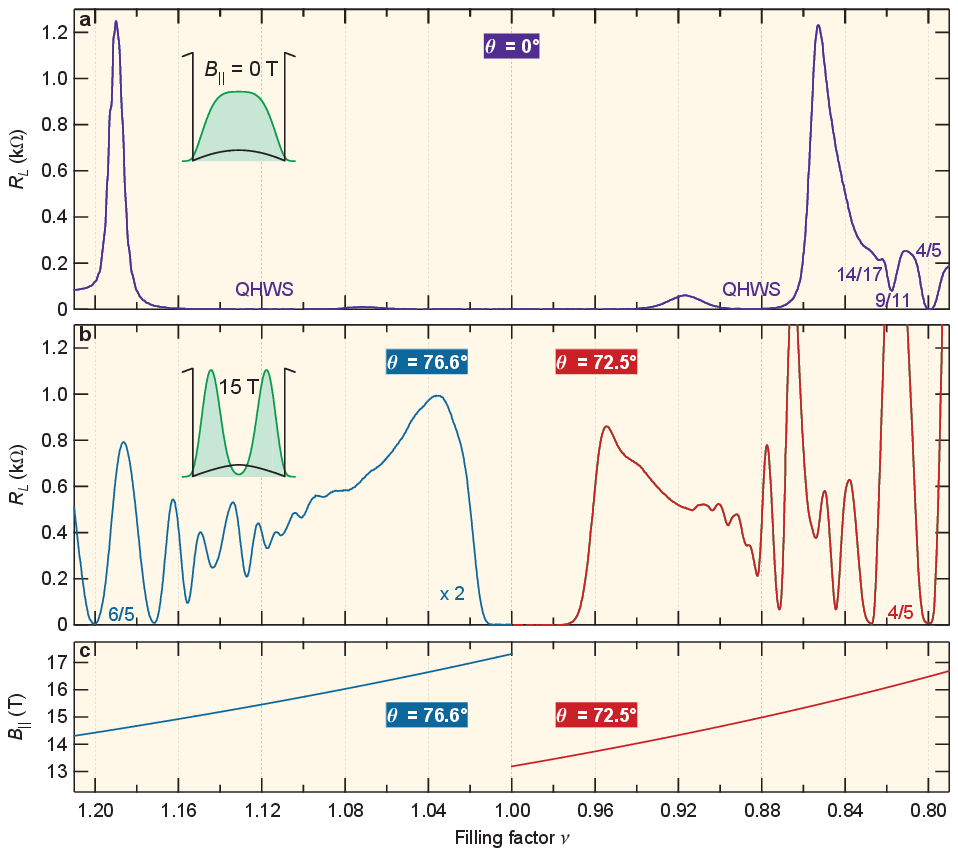}
\vspace{-0.1 in}
\caption{\small{
Longitudinal resistance $\rl$ in (a) perpendicular magnetic field $(\theta = 0^\circ)$ and (b) in tilted magnetic fields $(\theta = \tc, \nu > 1)$, $(\theta = \ta, \nu < 1)$ as a function of filling factor $\nu$.
Insets illustrate electron distribution within our quantum well at (a) $\bip = 0$ and (b) $\bip = 15$ T, a representative value, as illustrated in panel (c), showing $\bip$, corresponding to panel (b), versus $\nu$. 
}}
\label{fig1}
\vspace{-0.15 in}
\end{figure*}
In \rfig{fig1} we present the longitudinal resistance $\rl$ as a function of the filling factor $\nu$ at (a) $\theta = 0^\circ$ and at (b) $\theta = \tc$ ($\nu > 1$), $\theta = \ta$ ($\nu < 1$), covering the range $6/5 \lesssim \nu \lesssim 4/5$.
In a purely perpendicular magnetic field, see \rfig{fig1}\,(a), the data are dominated by a wide quantum Hall state at $\nu = 1$ and its re-entrant satellites, which we interpret as signatures of a quantum Hall Wigner solid (QHWS) \citep{huang:2022b,myers:2024}.
In addition, at $\nu < 1$ the data show the formation of FQH states at $\nu = 1 - 1/5 = 4/5$ and at $\nu = 1 - 2/11 = 9/11$ \cite{huang:2024}, both of which can be understood in terms of six-flux CFs, see \req{eq.jain} for $2p = 6$.
There is also a weak minimum, which likely signals an onset of the next six-flux CF state, at $\nu = 1 - 3/17 = 14/17$.
Indeed, all of these states are of a single-component origin.
In contrast, we see no signatures of corresponding particle-hole counterparts of these states at $\nu > 1$, which would be expected at $\nu = 6/5, 13/11$, and $20/17$, respectively.
The QHWS states are also asymmetric with respect to $\nu = 1$; they have different widths, appearing at $1.08 \lesssim \nu \lesssim 1.16$ on one side and at $0.87 \lesssim \nu \lesssim 0.90$ on the other.
These observations suggest that the particle-hole symmetry is not respected at $\theta = 0$.

Intriguingly, the relatively simple $\rl$ landscape seen in\rfig{fig1}\,(a)  changes dramatically upon tilting the sample, see \rfig{fig1}\,(b).
While the state at $\nu = 1$ remains quantized, its width is reduced dramatically, and the insulating QHWS states disappear entirely.
Instead, we observe many $\rl$ minima which persist up to $|\nu  - 1| \lesssim 0.08$.
As we show below, these minima correspond to 2C-FQH states, symmetrically flanking the $\nu = 1$ quantum Hall state.
We will also demonstrate that these new 2C-FQH states obey particle-hole symmetry, as can already be inferred from states at $\nu^{+}_{22} = 4/5$ and at $\nu^-_{33} = 6/5$.

\begin{figure*}[t] 
\includegraphics{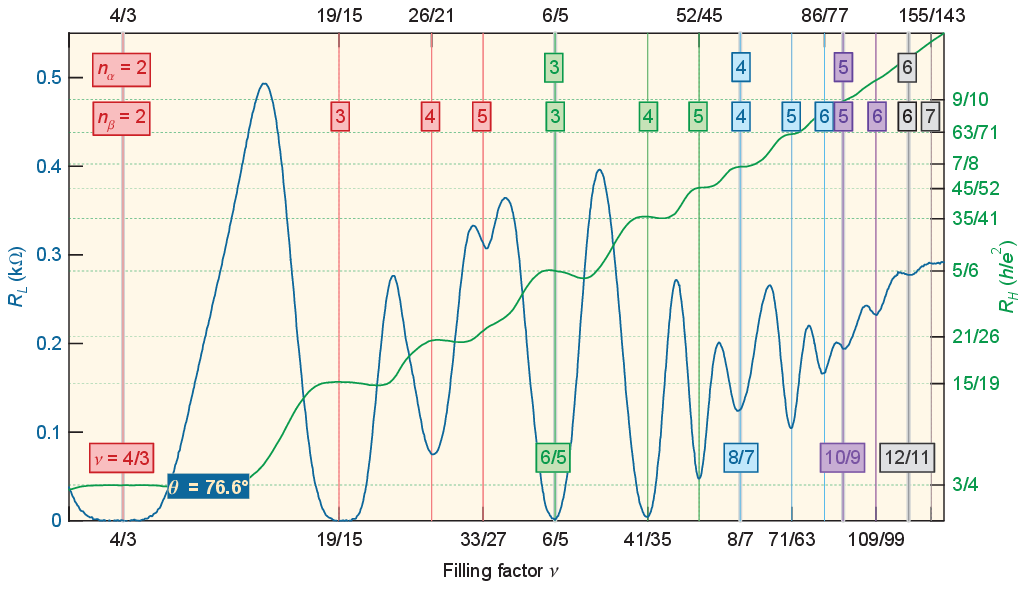}
\vspace{-0.1 in}
\caption{\small{
Longitudinal resistance $\rl$ (left axis, in k$\Omega$) and Hall resistance $\rh$ (right axis, in units of von Klitzing constant, $R_K = h/e^2$) as a function of $\nu$ 
measured at $\theta = \tc$.
Thick vertical lines mark balanced (parent) 2C-FQH states and thin vertical lines are drawn at $\nu$ corresponding to the unbalanced (daughter) states. 
Near the top axis we also mark the CF filling factors in each layer, $n_\alpha$ and $n_\beta$.
}}
\label{fig2}
\vspace{-0.15 in}
\end{figure*}
To illustrate the two-component nature of these new FQH states, we will first focus on the $\nu > 1$ range, extending it to $\nu \approx 4/3$.
In \rfig{fig2}, we show both the longitudinal resistance $\rl$ (left axis, in k$\Omega$) and the Hall resistance $\rh$ (right axis, in units of the von Klitzing constant, $R_K = h/e^2$) as a function of $\nu$, measured at the same angle as in \rfig{fig1}\,(b),  $\theta = \tc$.
The vertical lines are drawn at filling factors $\nu^-_{n_\alpha n_\beta}$ computed using \req{eq.2c} (also shown in Table\,I) and are marked on both the bottom and the top axis (to avoid the overlap).
The values of CF filling factors, $n_\alpha$ and $n_\beta \ge n_\alpha$, are shown near the top axis.
We indeed see that calculated lines nicely match the positions of the experimentally observed $\rl$ minima.

\begin{table}[b]
\vspace{-0.1in}
  \centering
\begin{tabularx}{\columnwidth} {lllllllllllll}
\hline  \hline \\[-2.2ex] 
    $n_\alpha$$n_\beta$ & $\nu^+_\alpha$ & $\nu^+_\beta$ & $\nu^+_{n_\alpha n_\beta}$ & $\delta \tilde \nu^+$  & $n_\alpha$$n_\beta$ & $\nu^-_\alpha$ & $\nu^-_\beta$ & $\nu^-_{n_\alpha n_\beta}$ & $\delta \tilde \nu^-$\\
    \hline \\ [-2ex]
    \textbf{1 1} & 1/3 & 1/3   & 2/3       & 0     & \textbf{2 2}  & 2/3 & 2/3 & 4/3 & 0 \\
     1 2 & 1/3 & 2/5   & 11/15     & 1/11  &  2 3 & 2/3 & 3/5 & 19/15 & 1/19 \\
       1 3 & 1/3 & 3/7   & 16/21     & 1/8   &  2 4 & 2/3 & 4/7 & 26/21 & 1/13 \\
       1 4 & 1/3 & 4/9   & 21/27       & 1/7   &  2 5 & 2/3 & 5/9 & 11/9 & 1/11 \\
     \hline \\ [-2ex]
    \textbf{2 2} & 2/5 & 2/5   & 4/5       & 0     & \textbf{3 3}  & 3/5 & 3/5 & 6/5 & 0 \\
      2 3 & 2/5 & 3/7   & 29/35     & 1/29  &  3 4 & 3/5 & 4/7 & 41/35 & 1/41 \\
      2 4 & 2/5 & 4/9   & 38/45     & 1/19  &  3 5 & 3/5 & 5/9 & 52/45 & 1/26 \\
    \hline \\ [-2ex]
    \textbf{3 3} & 3/7 & 3/7   & 6/7       & 0     & \textbf{4 4} & 4/7 & 4/7 & 8/7 & 0 \\  
     3 4 & 3/7 & 4/9   & 55/63     & 1/55  &  4 5 & 4/7 & 5/9 & 71/63 & 1/71 \\
     3 5 & 3/7 & 5/11  & 68/77     & 1/34  &  4 6 & 4/7 & 6/11 & 86/77 & 1/43 \\
    \hline \\ [-2ex]
    \textbf{4 4} & 4/9 & 4/9   & 8/9       & 0     & \textbf{5 5} & 5/9 & 5/9 & 10/9 & 0 \\ 
     4 5 & 4/9 & 5/11  & 89/99     & 1/89  &  5 6 & 5/9 & 6/11 & 109/99 & 1/109 \\
    \hline \\ [-2ex]
    \textbf{5 5} & 5/11 & 5/11 & 10/11     & 0     & \textbf{6 6} & 6/11 & 6/11 & 12/11 & 0 \\ 
     5 6 & 5/11 & 6/13 & 131/143   & 1/131 &  6 7 & 6/11 & 7/13 & 155/143 & 1/155 \\ 
\hline  \hline 
  \end{tabularx}
\vspace{-0.05in}
\caption{\small{
CF filling factor pairs $n_\alpha$$n_\beta$, layer filling factors $\nu^\pm_\alpha, \nu^\pm_\beta$, total filling factors $\nu^\pm_{n_\alpha n_\beta} = \nu^\pm_\alpha + \nu^\pm_\beta$, and relative interlayer imbalance $\delta \tilde \nu^\pm = (\nu^\pm_\beta - \nu^\pm_\alpha)/(\nu^\pm_\beta + \nu^\pm_\alpha)$. 
Balanced (parent) 2C-FQH states are shown in bold face. 
Each row shows a $\nu^+_{n_\alpha n_\beta}$ state and its particle-hole symmetric counterpart at  $\nu^-_{n_\alpha n_\beta} = 2 - \nu^+_{n_\alpha n_\beta}$.
\vspace{-0.15in}
}}
\end{table}

\begin{figure}[hb] 
\vspace{-0.1 in}
\includegraphics{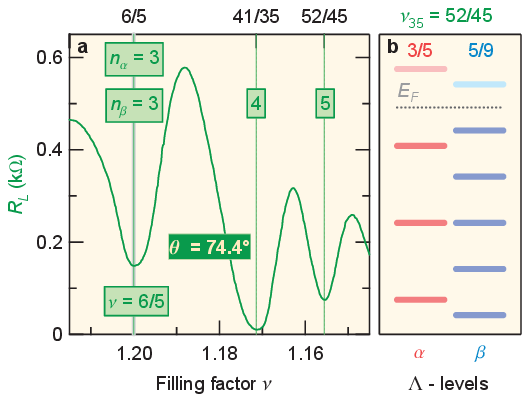}
\vspace{-0.25 in}
\caption{\small{
(a) Longitudinal resistance $\rl$ as a function of $\nu$, measured at $\theta = \td$, showing the $\nu^-_{3n_\beta}$ family of states, $\nu^-_{33} =6/5$, $\nu^-_{34} = 41/35$, and $\nu^-_{35} = 52/45$.
The CF filling factors in each layer, $n_\alpha$ and $n_\beta$, are marked near the top axis.
(b) $\Lambda$-levels in layers $\alpha$ and $\beta$ at $\nu^-_{35} = 52/45$ ($\nu^-_\alpha = 3/5, \nu^-_\beta = 5/9)$.
Levels below (above) the Fermi level $E_F$ (dotted line) are full (empty).
}}
\label{fig3}
\vspace{-0.15 in}
\end{figure}

\begin{figure*}[ht] 
\includegraphics{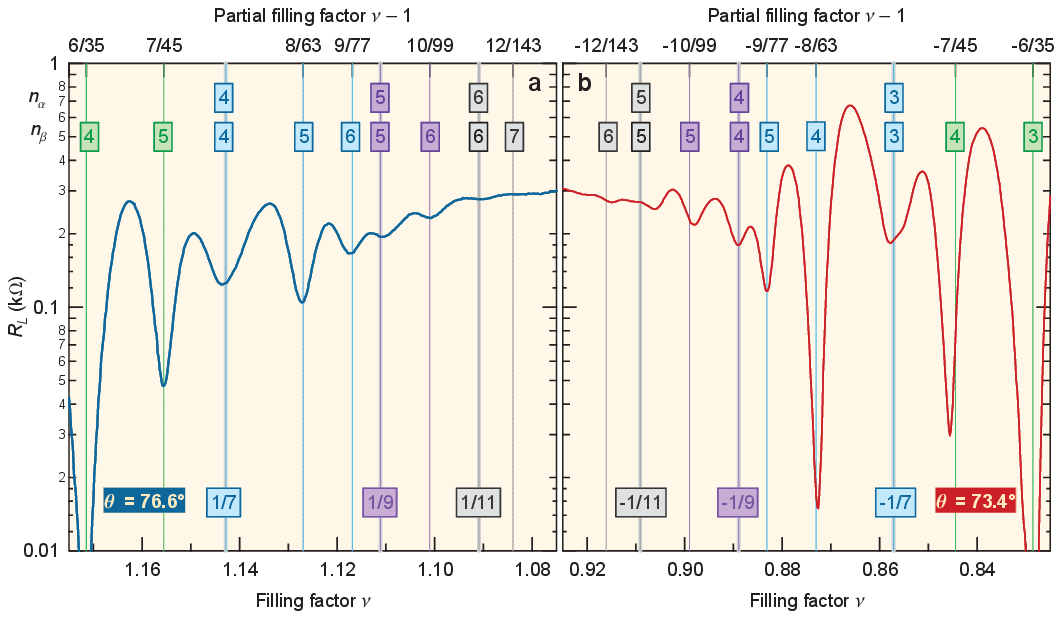}
\vspace{-0.1 in}
\caption{\small{
Longitudinal resistance $\rl$ as a function of filling factor $\nu$ (bottom axis) and partial filling factor $\tilde\nu = \nu - 1$ (top axis), measured at (a) $\theta = \tc$ and (b) $\theta = \tb$.
As in \rfig{fig2}, thick vertical lines mark balanced (parent) 2C-FQH states and thin lines mark unbalanced (daughter) states. 
The CF filling factors, $n_\alpha$ and $n_\beta$ and partial fillings of parent states are shown near the top and bottom axes, respectively. 
}}
\label{fig4}
\vspace{-0.15 in}
\end{figure*}

In a symmetric bilayer system, as mentioned above, one naturally expects that 2C-FQH states at double Jain's fractions would be found.
We indeed observe such states at  $\nu_{22}^- = 4/3$, $\nu_{33}^- = 6/5$, $\nu_{44}^- = 8/7$, $\nu_{55}^- = 10/9$, and $\nu_{66}^- = 12/11$, which are marked by thick vertical lines and labels near the bottom axis in \rfig{fig2}.
These fractions are sums of equal Jain fractions in each layer, $\nua^- = \nub^- = n/(2n-1) = 2/3,3/5,4/7,5/9$, and $6/11$.
In what follows we will refer to these simplest 2C-FQH states as either {\em balanced} ($\nua = \nub$) or {\em parent} states.

Apart from balanced  states at $\nu_{22}^- = 4/3$ and $\nu_{33}^- = 6/5$, the other well developed states are those at $\nu_{23}^- = 19/15$ and at $\nu_{34}^- = 41/35$, which is a particle-hole symmetric state of a $\nu_{23}^+ = 29/35$ state earlier reported \cite{manoharan:1997}.
These unbalanced $\nu_{23}^- =19/15$ and $\nu_{34}^- = 41/35$ states are the first {\em daughter} states of their respective parent states, $\nu_{22}^- = 4/3$ and $\nu_{33}^- = 6/5$, see Table\,I.
In fact, for a family of states originating from $\nu_{22}^- = 4/3$ we can identify three daughter states, namely $\nu_{23}^- = 19/15$, $\nu_{24}^- = 26/21$, and $\nu_{25}^- = 33/27$.
However, the parent state at $\nu_{33} = 6/5$ has only two daughter states, $\nu^-_{34} = 41/35$ and $\nu^-_{35} = 52/45$.
While the next parent state at $\nu^-_{44} = 8/7$ still has two daughter states, it is clear that the number of daughter states declines with increasing $n_\alpha$. 
This finding is expected since the range of $\nu$ in which daughter states might develop decreases as $\nu$ approaches unity.

Since only a few of the simplest daughter states have been observed until now, the emergence of high-order fractions, such as $\nu^-_{56} = 109/99$, may seem highly unlikely. 
However, this state is composed of relatively simple components, $\nu^-_\alpha =  5/9$ and $\nu^-_\beta  = 6/11$, which are indeed routinely seen in electron monolayers.
Indeed, $\rl$ measured in a purely perpendicular magnetic field shows signatures of FQH states extending up to much higher orders, $\nu^+ = 17/35$ and $\nu^- = 18/35$.
In addition, the imbalance of the $\nu^-_{56} = 109/99$ state is very small, less than 1\,\%, see Table\,I.

One can also expect that, for a given parent CF filling factor $n_\alpha$, the strength of daughter states would decrease monotonically with $n_\beta$ because the Jain's fractions involved in forming daughter states get weaker.
Indeed, as illustrated in \rfig{fig3}\,(b), the $\Lambda$-level spacing for $\nu_\beta = 5/9$ is smaller than for $\nu_\alpha = 3/5$ and should be the one which governs the gap of the $\nu^-_{35} = 52/45$ state.
One thus anticipates that parent states should always be stronger than their daughter states.
While this rule seems to hold for $\nu_{2n_\beta}^-$ and $\nu_{3n_\beta}^-$ families, it is clearly violated for higher-order families.
For example, the state at $\nu_{45}^- = 71/63$ is noticeably more robust than its parent $\nu_{44}^- = 8/7$ state, and the  same can be said about the $\nu_{56}^- = 109/99$ and $\nu_{55}^- = 10/9$ states.
In the data at $\theta = \td$, presented in \rfig{fig3}\,(a), we also see that the $\nu^-_{34} = 41/35$ state is stronger than its parent $\nu^-_{33} = 6/5$ state.
The reason for such behavior is unclear at this point, but it might be related to a slight asymmetry of the quantum well.
Nevertheless, this unusual trend seems to be generic as it is also observed at $\nu < 1$.

The $R_L$ minima seen in \rfig{fig1} at $\nu < 1$ can be explained in a similar way, but with $\nu^+_{n_\alpha n_\beta}$.
Here, the parent states are $\nu_{11}^+ = 2/3$, $\nu_{22}^+ = 4/5$, $\nu_{33}^+ = 6/7$,...\,, which are particle-hole symmetric counterparts of $\nu_{22}^- = 4/3$, $\nu_{33}^- = 6/5$, $\nu_{44}^- = 8/7$,...\,, respectively.
The daughter states can then be constructed using $\nu_{n_\alpha n_\beta}^+ = 2 - \nu_{(n_\alpha+1)(n_\beta+1)}^-$, see Table\,I.
To illustrate the particle-hole symmetry about $\nu = 1$, we introduce a partial filling factor $\tilde\nu = \nu - 1$.
With this notation, e.g., $\nu_{44}^- = 8/7$ and $\nu_{33}^+ = 6/7$  map to ``electron'' partial filling factor $\tilde\nu = 1/7$ and ``hole'' partial filling factor $\tilde\nu = -1/7$, respectively. 
In \rfig{fig4}, we compare $R_L$ measured at (a) $\nu > 1$ ($\theta = \tc$) and $R_L$ measured at (b) $\nu < 1$  ($\theta = \tb$), and displayed as a function of both $\nu$ (bottom axes) and $\tilde\nu$ (top axes).
As in \rfig{fig2}, expected 2C-FQH states are labeled by $n_\alpha$ and $n_\beta$ near the top axes.
The data clearly reflect particle-hole symmetry around $\nu = 1$.
We further observe that the $\nu^+_{34} = 55/63$ state, a particle-hole counterpart of the $\nu_{45}^- = 71/63$ state, is considerably stronger than its parent state $\nu^+_{33} = 6/7$.
Similarly, the state at $\nu_{45}^+ = 89/99$ state is more robust than that at $\nu_{44}^+ = 8/9$, consistent with their respective counterparts at $\nu_{56}^- = 109/99$ and $\nu_{55}^- = 10/9$.
We thus conclude that the 2C-FQH states obey the particle-hole symmetry but not necessarily the expected Jain hierarchy.

In summary, we have experimentally demonstrated the existence of several families hosting high-order two-component fractional quantum Hall (2C-FQH) states. 
These states emerge symmetrically around $\nu = 1$, respect particle-hole symmetry, and extend to CF filling factor differences as high as three. 
A particularly surprising finding is that many daughter states are stronger than their parent states, which is at odds with conventional expectations and suggests that correlations play a more central role in stabilizing bilayer FQH phases than previously thought. 
The discovery of these higher-order 2C-FQH phases broadens the landscape of correlated quantum matter and provides fresh insight into the interplay of symmetry, interactions, and topology in two-dimensional electron systems.
Future work should aim to identify the mechanisms responsible for the enhanced strength of unbalanced states, explore whether similar families of states can be realized in other material platform, and search for inter-layer-correlated states \citep{scarola:2001}, as found in graphene bilayers \citep{li:2019}. 

\begin{acknowledgments}
We thank Y. Huang, J. Jain, Y. Liu, Q. Shi, and B. Shklovskii for discussions, and  A. Bangura, E. Choi, S. Hannas, G. Jones, L. Jiao, T. Murphy, R. Nowell, and  A. Woods for technical support.
Growth of GaAs/AlGaAs quantum wells at Princeton University was supported in part by the Gordon and Betty Moore Foundation’s EPiQS Initiative, Grant GBMF9615 to L. N. Pfeiffer, and by the National Science Foundation MRSEC grant DMR 1420541. 
A portion of this work was performed at the National High Magnetic Field Laboratory, which is supported by National Science Foundation Cooperative Agreements No. DMR-1644779, No. DMR-2128556, and the State of Florida.
\end{acknowledgments}
%

\end{document}